\documentclass[11pt,twoside]{article}

\usepackage{asp2014}

\aspSuppressVolSlug
\resetcounters

\begin{document}

\title{Container-Based Pre-Pipeline Data Processing on HPC for XRISM}

\author{Satoshi~Eguchi,$^{1}$
  Makoto~Tashiro,$^{2,3}$
  Yukikatsu~Terada,$^{2,3}$
  Hiromitsu~Takahashi,$^{4}$
  Masayoshi~Nobukawa,$^{5}$
  Ken~Ebisawa,$^{3}$
  Katsuhiro~Hayashi,$^{3}$
  Tessei~Yoshida,$^{3}$
  Yoshiaki~Kanemaru,$^{3}$
  Shoji~Ogawa,$^{3}$
  Matthew~P.~Holland,$^{6}$
  Michael~Loewenstein,$^{7,6,8}$
  Eric~D.~Miller,$^{9}$
  Tahir~Yaqoob,$^{10,6,8}$
  Robert~S.~Hill,$^{11,6}$
  Morgan~D.~Waddy,$^{11,6}$
  Mark~M.~Mekosh,$^{11,6}$
  Joseph~B.~Fox,$^{11,6}$
  Isabella~S.~Brewer,$^{11,6}$
  Emily~Aldoretta,$^{11,6}$
  and XRISM Science Operations Team}
\affil{$^{1}$Kumamoto Gakuen University, Kumamoto, Japan; \email{sa-eguchi@kumagaku.ac.jp}}
\affil{$^{2}$Saitama University, Saitama, Japan}
\affil{$^{3}$Institute of Space and Astronautical Science, Japan Aerospace Exploration Agency, Sagamihara, Japan}
\affil{$^{4}$Hiroshima University, Higashi-Hiroshima, Japan}
\affil{$^{5}$Nara University of Education, Nara, Japan}
\affil{$^{6}$National Aeronautics and Space Administration, Goddard Space Flight Center, Greenbelt, Maryland, United States}
\affil{$^{7}$University of Maryland, College Park, Maryland, United States}
\affil{$^{8}$Center for Research and Exploration in Space Science and Technology (CRESST), Greenbelt, Maryland, United States}
\affil{$^{9}$Massachusetts Institute of Technology, Cambridge, Massachusetts, United States}
\affil{$^{10}$University of Maryland, Baltimore County, Baltimore, Maryland, United States}
\affil{$^{11}$ADNET Systems, Inc., Bethesda, Maryland, United States}

\paperauthor{Satoshi~Eguchi}{sa-eguchi@kumagaku.ac.jp}{0000-0003-2814-9336}{Kumamoto Gakuen University}{Department of Economics}{Kumamoto}{Kumamoto}{862-8680}{Japan}
\paperauthor{Makoto~Tashiro}{tashiro@mail.saitama-u.ac.jp}{}{Saitama University}{Graduate School of Science and Engineering}{Saitama}{Saitama}{338-8570}{Japan}
\paperauthor{Yukikatsu~Terada}{terada@mail.saitama-u.ac.jp}{}{Saitama University}{Graduate School of Science and Engineering}{Saitama}{Saitama}{338-8570}{Japan}
\paperauthor{Hiromitsu~Takahashi}{hirotaka@astro.hiroshima-u.ac.jp}{}{Hiroshima University}{School of Science}{Higashi-Hiroshima}{Hiroshima}{739-8511}{Japan}
\paperauthor{Masayoshi~Nobukawa}{nobukawa@cc.nara-edu.ac.jp}{}{Nara University of Education}{Department of Teacher Training and School Education}{Nara}{Nara}{630-8528}{Japan}
\paperauthor{Ken~Ebisawa}{ebisawa.ken@jaxa.jp}{}{Institute of Space and Astronautical Science, Japan Aerospace Exploration Agency}{}{Sagamihara}{Kanagawa}{252-5210}{Japan}
\paperauthor{Katsuhiro~Hayashi}{hayashi.katsuhiro@jaxa.jp}{}{Institute of Space and Astronautical Science, Japan Aerospace Exploration Agency}{}{Sagamihara}{Kanagawa}{252-5210}{Japan}
\paperauthor{Tessei~Yoshida}{yoshida.tessei@jaxa.jp}{}{Institute of Space and Astronautical Science, Japan Aerospace Exploration Agency}{}{Sagamihara}{Kanagawa}{252-5210}{Japan}
\paperauthor{Yoshiaki~Kanemaru}{kanemaru.yoshiaki@jaxa.jp}{}{Institute of Space and Astronautical Science, Japan Aerospace Exploration Agency}{}{Sagamihara}{Kanagawa}{252-5210}{Japan}
\paperauthor{Shoji~Ogawa}{ogawa.shohji@jaxa.jp}{}{Institute of Space and Astronautical Science, Japan Aerospace Exploration Agency}{}{Sagamihara}{Kanagawa}{252-5210}{Japan}
\paperauthor{Matthew~P.~Holland}{matthew.p.holland@nasa.gov}{}{National Aeronautics and Space Administration}{Goddard Space Flight Center}{Greenbelt}{Maryland}{20771}{United States}
\paperauthor{Michael~Loewenstein}{michael.loewenstein-1@nasa.gov}{}{University of Maryland}{}{College Park}{Maryland}{20742}{United States}
\paperauthor{Eric~D.~Miller}{milleric@mit.edu}{}{Massachusetts Institute of Technology}{MIT Kavli Institute for Astrophysics and Space Research}{Cambridge}{Massachusetts}{02139}{United States}
\paperauthor{Tahir~Yaqoob}{tahir.yaqoob-1@nasa.gov}{}{University of Maryland}{}{Baltimore}{Maryland}{21250}{United States}
\paperauthor{Robert~S.~Hill}{robert.s.hill@nasa.gov}{}{ADNET Systems, Inc.}{}{Bethesda}{Maryland}{20817}{United States}
\paperauthor{Morgan~D.~Waddy}{morgan.d.waddy@nasa.gov}{}{ADNET Systems, Inc.}{}{Bethesda}{Maryland}{20817}{United States}
\paperauthor{Mark~M.~Mekosh}{mark.m.mekosh@nasa.gov}{}{ADNET Systems, Inc.}{}{Bethesda}{Maryland}{20817}{United States}
\paperauthor{Joseph~B.~Fox}{joseph.b.fox@nasa.gov}{}{ADNET Systems, Inc.}{}{Bethesda}{Maryland}{20817}{United States}
\paperauthor{Isabella~S.~Brewer}{isabella.s.brewer@nasa.gov}{}{ADNET Systems, Inc.}{}{Bethesda}{Maryland}{20817}{United States}
\paperauthor{Emily~Aldoretta}{emily.aldoretta@nasa.gov}{}{ADNET Systems, Inc.}{}{Bethesda}{Maryland}{20817}{United States}
\paperauthor{XRISM Science Operations Team}{}{}{}{}{}{}{}{}



\begin{abstract}
The X-Ray Imaging and Spectroscopy Mission (XRISM) is the 7th Japanese X-ray observatory,
whose development and operation are in collaboration with universities and research institutes in Japan, U.S., and Europe, including JAXA, NASA, and ESA.
The telemetry data downlinked from the satellite are reduced to scientific products by
the pre-pipeline (PPL) and pipeline (PL) software running on standard Linux virtual machines
on the JAXA and NASA sides, respectively.
We ported the PPL to the JAXA ``TOKI-RURI'' high-performance computing (HPC) system capable
of completing $\simeq 160$ PPL processes within 24 hours by utilizing the container platform
of Singularity and its ``--bind'' option.
In this paper, we briefly show the data processing in XRISM and present our porting strategy of
PPL to the HPC environment in detail.
\end{abstract}



\section{Introduction}

The X-Ray Imaging and Spectroscopy Mission (XRISM) is the 7th Japanese X-ray observatory,
equipped with the revolutionary X-ray microcalorimeter array (Resolve) and the X-ray CCD camera (Xtend),
is expected to provide a leap in our understanding of the formation and evolution of the Universe,
galaxies, compact objects, and supernovae.
The project is led by the Japan Aerospace Exploration Agency (JAXA) and the National Aeronautics and
Space Administration (NASA) in collaboration with the European Space Agency (ESA) and other partners.
The Science Operations Center (SOC) at JAXA and the Science Data Center (SDC) at NASA are responsible
for data processing and distribution \citep{2021JATIS...7c7001T}.

\section{Data Processing in XRISM}

\articlefigure{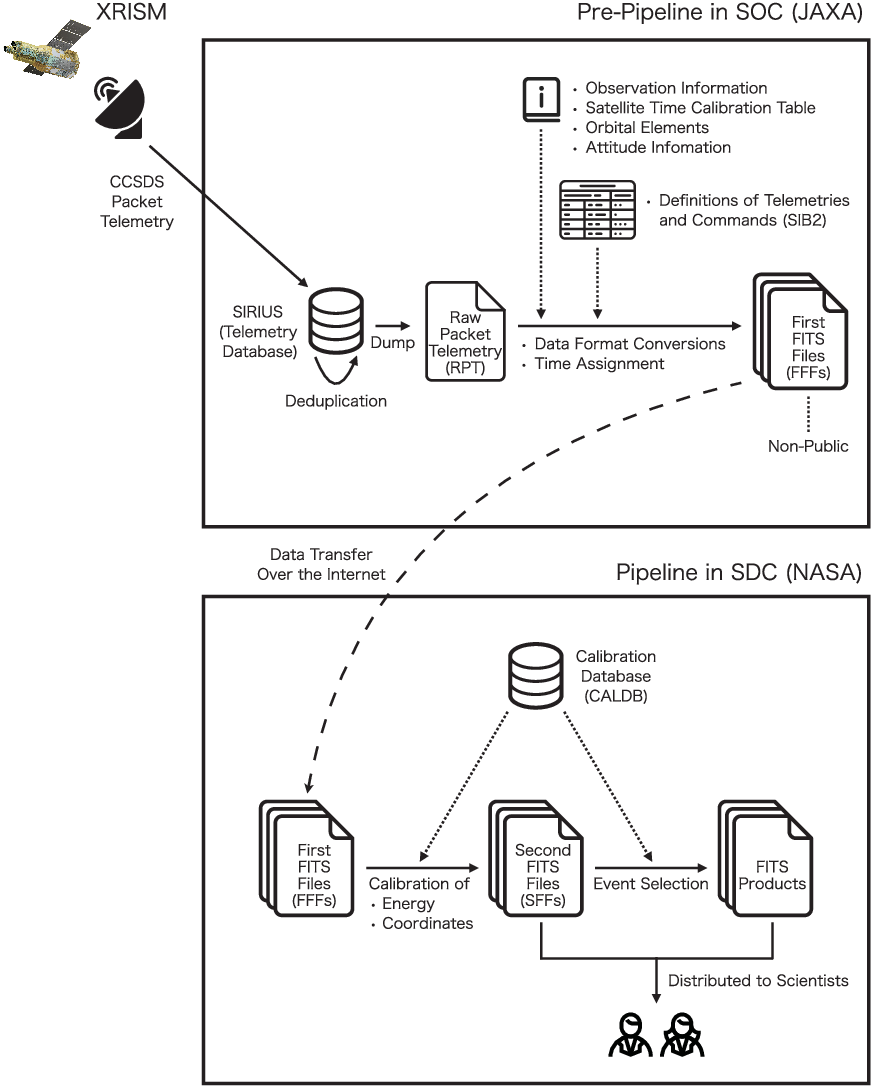}{fig1}{Schematic diagram of data processing in XRISM.
 Telemetry data downlinked from XRISM are converted into FITS products with the pre-pipeline
 and pipeline software in SOC and SDC, respectively, and distributed by both centers.\citep{Eguchi2025}}


In XRISM, observational data are reduced by the pre-pipeline software (PPL) in SOC and the pipeline
software (PL) in SDC.
Figure.~\ref{fig1} shows the schematic diagram of data reduction for pipeline products in XRISM.
The telemetry data downlinked from XRISM as CCSDS packets are stored in SIRIUS, the telemetry database
operated by C-SODA in JAXA, and duplicate packets are removed.
During the first stage of PPL, CCSDS packets are dumped into a FITS file named ``Raw Packet Telemetry'' (RPT).
The packets in the RPT are compiled into essential raw values based on the information of the satellite
and observation and stored in multiple ``First FITS Files'' (FFFs).
The FFFs are transferred to SDC over the Internet and reduced to the ``Second FITS Files'' (SFFs) and
FITS products (ready-for-analysis) by the PL.

\section{PPL Reprocessing on HPC}

PPL and PL are still under improvement, and whole (> 100) products must be reprocessed simultaneously
with the latest version of PPL and PL at some points to homogenize the quality since the PPL processing
cannot be applied by end users.
Both software runs on standard Linux virtual machines (VMs).
We focus on data reprocessing with PPL here.
C-SODA provides the software used for conversions between CCSDS packets and raw values as 32-bit executables
and is not aware of 64-bit inode numbers (this is the case for XFS, for instance);
this made us not simply port PPL to the JAXA ``TOKI-RURI'' PC cluster but utilize virtualization technologies
that work on the high-performance computing (HPC) system.
We found that Singularity, a container platform, Ver. 3.10 was available on the system.
On ``Reformatter,'' our VM for PPL, the directories required for PPL are managed by symbolic links, i.e.,
virtually hard-coded. On the other hand, we can create arbitrary mappings between a real file system and that
inside a container instance by using the ``\texttt{--bind}'' option of Singularity.
This option also accepts disk images formatted to ext3, whose inode numbers are within the 32-bit range.
Thus, we decided to utilize Singularity to ``replicate'' Reformatter, including the directory structure, on TOKI-RURI.

\articlefigure{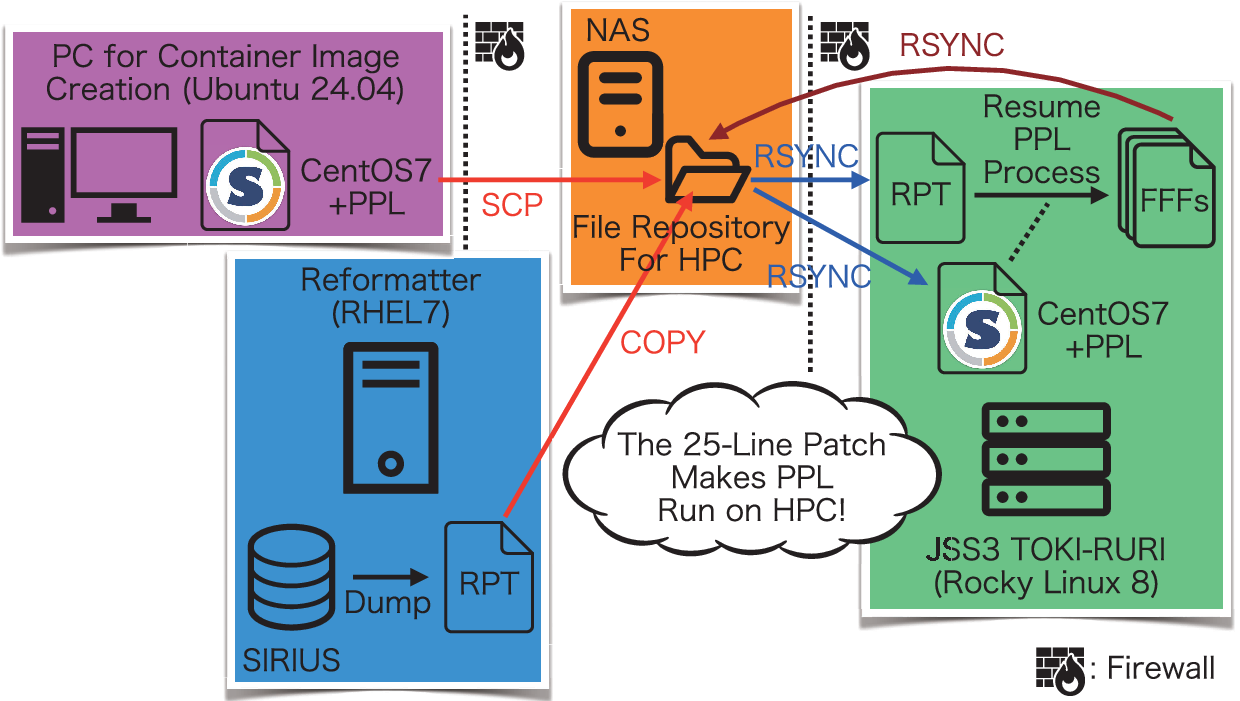}{fig2}{Schematic data flow of the HPC version of pre-pipeline.
  There are firewalls between a PC for container image creation, Reformatter, and TOKI-RURI.}

Figure.~\ref{fig2} shows the schematic data flow of the HPC version of PPL.
Since machines accessible to SIRIUS are limited, we create a set of RPT files required for reprocessing on Reformatter.
The RPTs are put in a file repository for data transfer to the HPC on our NAS.
A Singularity container image is created on a PC other than Reformatter and copied to the repository.
Once all necessary files are accumulated, the repository is archived into a tarball with Zstandard compression
and copied to the HPC with the \texttt{rsync} command, whose bandwidth is $\simeq 50$ MiB/s.
When the tarball is extracted on the HPC, a Python script generates a job script to resume the PPL tasks for each RPT.
The job script also attaches necessary working disk images to the Singularity instance.
The Python script automatically submits the job scripts.
The outputs are archived into a tarball and sent back to the NAS.

\section{Results and Conclusion}

We reprocessed all the observation data (identified by OBSIDs) with the HPC PPL in March and September 2024, respectively (see \citet{Eguchi2025} for details).
We focus on the latter in this paper.
There were 161 OBSIDs, and the total time required for the data reprocessing on Reformatter was estimated to be 515.5 hours,
of which $\simeq 3\%$ was expected due to RPT creations.
On the other hand, TOKI-RURI completed the tasks within 15.2 hours;
we obtain the speedup
\begin{displaymath}
 S = \dfrac{515.5 \cdot \left( 1 - 0.03 \right)}{15.2} = 33.
\end{displaymath}
Thus, we can conclude that we successfully ported the PPL to the HPC system and achieved sufficient speedup.

\acknowledgements This work was supported by JSPS Core-to-Core Program \\ (grant number: JPJSCCA20220002).

\bibliography{main}  


\end{document}